# Quantifying the Topology of Magnetic Skyrmions in three Dimensions

David Raftrey[1,2], Simone Finizio[3], Rajesh V. Chopdekar[4], Scott Dhuey[5], Temuujin Bayaraa[1], Paul Ashby[5], Jörg Raabe[3], Tiffany Santos[4], Sinéad Griffin[1,5], Peter Fischer[1,2]

[1]Materials Sciences Division, Lawrence Berkeley National Laboratory, Berkeley CA 94720, USA
[2]Physics Department, University of California Santa Cruz, Santa Cruz CA 95064, USA
[3]Swiss Light Source, Paul Scherrer Institute, 5232 Villigen PSI, Switzerland
[4]Western Digital Research Center, Western Digital Corporation, San Jose CA 95119, USA
[5]Molecular Foundry, Lawrence Berkeley National Laboratory, Berkeley CA 94720, USA

**Abstract**

Magnetic skyrmions have so far been treated as two-dimensional spin structures characterized by a topological winding number describing the rotation of spins across the skyrmion. However, in real systems with a finite thickness of the material being larger than the magnetic exchange length, the skyrmion spin texture extends into the third dimension and cannot be assumed as homogeneous. Using soft x-ray laminography we reconstruct with about 20nm spatial (voxel) resolution the full three-dimensional spin texture of a skyrmion in an 800 nm diameter and 95 nm thin disk patterned into a trilayer [Ir/Co/Pt] thin film structure. A quantitative analysis finds that the evolution of the radial profile of the topological skyrmion number and the chirality is non-uniform across the thickness of the disk. Estimates of local micromagnetic energy densities suggest that the changes in topological profile are related to non-uniform competing energetic interactions. Theoretical calculations and micromagnetic simulations are consistent with the experimental findings.

Our results provide the foundation for nanoscale magnetic metrology for future tailored spintronics devices using topology as a design parameter, and have the potential to reverse-engineer a spin Hamiltonian from macroscopic data, tying theory more closely to experiment.

**Main text**

Skyrmions – the essential ingredient in many next-generation spintronic applications – have conventionally been studied as two-dimensional objects owing to the inability to accurately characterize their properties into the third dimension. However, in real systems with a finite thickness larger than the magnetic exchange length, the details of the spin texture extending into the third dimension cannot be neglected, especially when considering their emergent properties and functionality. This is especially relevant when the topological properties are a feature of the extra dimensionality – for instance while rigid skyrmion tubes can readily be extended to 3D, more complex spin textures such as twisted skyrmion tubes [1, 2], Hopfions [3], torons, cocoons [4], and vortex rings [5] as well as ferroelectric polar skyrmions [6], and artificially designed magnetic nanostructures, such as braided wires, tetrapods etc. are among such 3D topological building blocks that require knowledge of the spin texture evolution in space[7, 8]. Advanced synthesis methods allow a highly precise engineering of materials that extend into the third dimension, and therefore a fundamental understanding of the full 3D spin texture opens opportunities to explore and tailor 3D topological spintronic devices with enhanced functionalities that cannot be achieved in two dimensions.

Magnetic skyrmions are stable, particle-like topological solitons found in specific magnetic materials. Visualized in two dimensions, they resemble circular structures where spins at the center of the skyrmion point in the opposite direction to those at their edges. Skyrmions can move within a material without changing their shape or size due to their topological protections. This makes them robust to defects and particularly promising for spintronic applications where information is stored and manipulated in the electron spin instead of the charge as in conventional electronics. This robustness is indexed by a topological charge: the skyrmion number or winding number which is defined as $N_{sk} = \iint d^2r \boldsymbol{m} \cdot (\frac{\partial \boldsymbol{m}}{\partial x} \times \frac{\partial \boldsymbol{m}}{\partial y})$ where $\boldsymbol{m}$ is the magnetization, and $N_{sk}$ counts how many times the direction of a spin wraps around a unit sphere, e.g. $N_{sk} = 1$ for a magnetic skyrmion.

The properties, behaviors and functionality of magnetic materials is determined and influenced by their microscopic magnetic textures, such as skyrmions. In particular, at a microscopic level such properties and functionality is determined by the energetic landscape with many competing interactions: the symmetric exchange (Heisenberg) interaction favors a parallel or antiparallel alignment of neighboring spins, while the antisymmetric exchange (Dzyaloshinskii-Moriya) interaction (DMI) favors non-collinear spin arrangements, both of which are short range, whereas the dipolar interaction is a long-range interaction. The magnetocrystalline anisotropy energy originates from the orbital moment, and the Zeeman energy is proportional to any external magnetic fields that the material experiences. Competition between these interactions determines the ground state magnetic order and the presence (or absence) of topologically protected spin textures.

A prerequisite for the development of future 3D spintronics is the availability of advanced characterization techniques that address energetic contributions and topology. Emerging approaches such as magnetic tomography techniques leverage advanced electron and x-ray-based microscopies. Among the most promising approaches are real space x-ray microscopies including full-field transmission soft x-ray microscopy, and diffraction-based x-ray techniques, like x-ray ptychography and x-ray laminography harness x-ray magnetic dichroism effects as strong, element-specific and, most notably quantitative, magnetic contrast to provide 3D images with a high level of details [9].

Here, we use soft x-ray laminography to capture the 3D spin texture of a magnetic skyrmion in a multilayered disk measuring 800 nm in diameter and 95 nm in thickness. Our analysis provides depth-dependent radial profiles of the skyrmion number and the chirality, laying the groundwork for nanoscale magnetic metrology. This technique can serve as a foundation for designing future tailored spintronics devices using topology as a design parameter, and opens up the field of 'magnetic tomography' for full three-dimensional reconstruction of bulk spin textures.

**- Depth profile of the topological skyrmion number and chirality**

The system used in this experiment is a magnetic multilayer of Ir/Co/Pt that was lithographically patterned into nanoscale disks. This class of multilayers is an established platform and is particularly appealing as it offers sensitive responses of the spin texture to changes and orderings of the thicknesses of the layers leading to topological textures such as skyrmions [10], target skyrmions [11], and Hopfions [3]. The origin of the topological textures in this system is the inversion symmetry breaking of the tri-layer stack which gives rise to a DMI interaction at the interface. In addition to DMI, the spin orbit coupling between the Co and Pt at the interface creates an (out-of-plane) perpendicular magnetic anisotropy (PMA) [12]. Previous studies have established that the DMI and PMA depend on layer thicknesses and ordering [13].

With this type of multilayer system (Ir/Co/Pt) and the ability to control the system through layer thicknesses requires to consider the third dimension. Particularly in the case of nano-patterned elements, as in this study the lateral dimension is on a comparable length scale than the height of the disk. Previously reported studies in such systems made with electron microscope imaging at a spatial resolution of several nanometers [2], and with hard x-rays on the micron length scale [5, 14] observed complex 3D spin textures. Element specific soft x-ray imaging with scanning transmission x-ray microscopy providing a spatial resolution in the tens of nanometer has the advantages of element specificity as compared to electron imaging as well as much faster acquisition time compared to hard x-rays due to the resonant enhancement of the circular dichroism in the vicinity of the $L_{3/2}$ x-ray absorption edges, which is particular advantageous for tomographic time-resolved experiments [15]. The combination of multilayer engineering, structural imprinting and soft x-ray laminographic imaging allows us to probe the microstructure at a length scale relevant for deducing micromagnetic parameters and energy densities within the spin textures with implication for nucleation and stability.

With only one single projection axis it is not possible to quantify the topological charge directly from the data. Typically, studies rely on theory-laden interpretation of data, often through indirect comparison between experimental data and projections of micromagnetic simulations.

On theoretical grounds, the classification of n-dimensional topological structures requires knowledge of all n-dimensions in the system [16]. Even for relatively thin systems, a two-dimensional approximation does not capture important aspects as chirality is typically not conserved along the depth profile with flux-closure structures at the top and bottom of the film, e.g. in Néel caps. Even several atomic layer differences in magnetic thin films may change chirality and topology [17]. For intrinsically three-dimensional structures such as Hopfions [3] or dipolar skyrmions [18], the topological features can only be quantitatively probed by means of a three-dimensional dataset. Several approaches have emerged with the capability of imaging topological magnetic structures at the resolution relevant to micromagnetic interactions and quantitative analysis of topological features including electron holography, x-ray ptychography, and as in this work, soft x-ray laminography [5, 19].

Importantly, we find that the profile of the topological charge across the disk has a significant and characteristic depth dependence (Fig. 3 a). The three-dimensional data allows us to quantitatively compute the skyrmion number or topological charge directly from the data (Fig. 3 a) in a discrete form. We find that, consistent with theory, the topological charge converges to a near-integer value when integrated radially. This value converges for all depth dependent layers. From the data it is possible to produce a three-dimensional rendering of the topological charge density which is proportional to the fringing field of the spin texture. This fringing field is found to be largest at the position of the domain wall of the skyrmion where the rotation of the texture leads to non-collinear spins creating uncompensated volume charges (fig. 3 a).

A micromagnetic curve fitting approach was used to analyze the energetics of the system. The quantity of interest for this system is the topological charge density $\rho_{sk}$ which is the integrated spatial quantity in the skyrmion number integral $\rho_{sk} = m \cdot (\frac{\partial m}{\partial x} \times \frac{\partial m}{\partial y})$. The spin texture arises from the competition between different energetic interactions. By radially integrating the topological charge density in the disk, a topological profile across the depth can be compared to micromagnetic simulations to estimate the variation of the relative energetic contributions from the different interactions i.e., PMA and DMI. (fig. 3 b, c). Particularly important for topological quantities is the DMI interaction which prefers non-collinear spins. The profile of the integrated topological charge density is fitted with the function $\frac{1}{(e^{-\alpha*(r+\rho)}+1)}$, with fitting parameters $\alpha$ and $\rho$ where $\alpha$ describes the profile shape and $\rho$ is a shift in radial position, $r$. We find from micromagnetic simulations that the fitting parameter $\alpha$, closely related to the skyrmion radius decreases with increasing DMI and increases with increasing anisotropy (Fig. 3 b, c). A larger skyrmion with a broader domain wall will have more spins canted which is preferred by the DMI. A smaller skyrmion with a thinner domain wall has the spins more perpendicular, following the out of plane anisotropy. The topological profile provides an especially well-defined curve for micromagnetic fitting procedures investigating the DMI vs PMA contribution. Using this curve, the depth dependence of energetics can be estimated. From the reconstruction, we find that the fitting parameter $\alpha$ is smallest for the top of the structure and also smallest at the bottom (Fig. 3 b) suggesting a barrel-like profile similar to the depth profile of a vortex core [20]. This suggests that the DMI and PMA vary across the stack (Fig. 3 c), which may be due to the changing interlayer roughness of successive layer depositions.

Chirality is another topological index that describes the rotation direction of the spins [21]. The chirality quantitatively describes if the domain wall is of Bloch or Néel type. The Bloch/Néel type is determined by crystal structure, the sign of the chirality is determined by details of the spin orbit interaction. Chirality is defined as the angle between the normal to the domain wall and the direction of the in-plane magnetization.

The chirality is quantified from data by first computing the gradient of the z component of the magnetization to find the normal vector to the domain wall, which lies in-plane. The angle between the normal vector and the in-plane components of magnetization is the chirality β. In the case of a Néel profile created by interfacial DMI, the angle is 180° in the case of positive chirality and 0° in the case of negative chirality. In the case of bulk type DMI like in B20 compounds the chirality is Bloch type with an angle of 90° or 270°. From the dataset, the chirality angle is computed for each voxel. The angles are binned and plotted in a histogram. To determine the depth dependence, the data are plotted with one line for each depth resolved layer. In this system, the chirality is constantly maximum at 180°, indicating a positive Néel profile across the depth (Fig, 3 d). The consistent Néel profile is consistent with theoretical predictions for a multilayer system.

- **Depth dependence of PMA, exchange interaction and DMI**

A three-dimensional measurement makes computation of micromagnetic energy densities possible. Unlike in a 2D projection where 3D information can only be inferred, with a direct 3D measurement vector quantities relating to terms in the micromagnetic Hamiltonian can be computed from data. Here we compare the contributions of competing magnetic interactions namely symmetric exchange $-A(\nabla \cdot m)^2$, antisymmetric exchange or Dzyaloshinskii Moriya Interaction (DMI) $-\vec{D} \cdot (m_i \times m_j)$, and uniaxial anisotropy $-K(\hat{z} \cdot m)^2$ [12, 13]. While this technique cannot measure the constants A, D and K directly, the values of vector operations on the data can be reported allowing us to spatially map the different energy densities.

Taking vector operations on the data shows the domain wall to be the region of maximal DMI and PMA while exchange energy is more localized to the core of the skyrmion (Fig. 4 a, b, c). At the core of the skyrmion, the divergence of the vector field is largest, leading to a maximum Heisenberg exchange contribution. In the domain wall, the in-plane spins are unfavorably aligned with the easy out of plane axis leading to a large accumulation of PMA energy. The DMI however lowers the energy of the domain wall as the vector $\vec{D}$ is in-plane for this multilayer system making energetic contributions to the azimuthally symmetric spins canting across the domain wall.

Along the z-profile, the calculated second-principals energy densities are non-uniform (Fig. 4 d, e, f). This suggests that the fundamental energetic interactions are non-uniform across the depth profile, which may be related to atomic scale roughness or pinning defects. The emergent consequence of the changing energetic densities is that the skyrmion profile evolves with depth (Fig. 3a). However, the change in local energy density does not appear to affect the chirality of the skyrmion (Fig. 3d).

With this technique the energy can only be determined at the level of spatial distribution. Knowledge of the magnetization field alone is not enough to directly determine values of micromagnetic constants, which are determined by both spin and orbital ordering on atomic length scales. Estimating energetic properties from direct vector operations cannot determine behavior below the ~20 nm resolution where interlayer coupling is relevant, and it cannot account for non-linear interactions between energetic terms.

Current theoretical models for skyrmions grapple with a handful of limitations that can restrict their precision and applicability. For one, many of these models operate on the assumption of homogeneity across the material. This can neglect important local variances in parameters like exchange interactions or anisotropy, which in reality, may fluctuate due to imperfections and inhomogeneities in the material. Moreover, while modeling of skyrmions as 2D structures is more straightforward, real-world systems extend into the third dimension such as complex spin textures in thin films or multilayers – in such systems the 2D assumption can lead to inaccuracies. In addition, most models assume static conditions, while in reality, skyrmions are dynamic entities. skyrmions can interact with their surroundings, move, and even change size. Lastly, though ab initio calculations serve as a robust foundation, they often lack experimental validation and can lead to discrepancies between theory and experiment.

The technique demonstrated here overcomes many of these limitations, without the need for computationally expensive (or intractable) calculations. Firstly, it embraces the inhomogeneity inherent in real-world systems by utilizing advanced imaging techniques, which capture variances in the interactions such as DMI strength across the sample. This valuable data is then incorporated into the micromagnetic models, enhancing their precision. Such information would not be accessible in *ab initio* calculations which typically approximate these complex systems in their ideal, crystalline form without defects. By analyzing depth-dependent radial profiles, it provides insight into 3D skyrmion dynamics.

Future skyrmion research and spintronics stand to gain from incorporating magnetic Hamiltonian tomography into theory development. By employing techniques such as soft x-ray laminography, full-field transmission soft x-ray microscopy, and other x-ray-based techniques, researchers can gain a comprehensive view of the 3D spin textures in magnetic materials. These techniques provide a more granular and accurate look at the magnetic textures, allowing for the creation of more precise theoretical models. Moreover, with these techniques, one could envision real-time tracking of skyrmion dynamics in 3D, offering deeper insights into their behavior. The combination of such imaging techniques with theoretical models will pave the way for next generation of spintronic devices, employing complex 3D topological building blocks for enhanced functionality.

To conclude, using magnetic soft x-ray laminography, we have experimentally obtained the 3D structure of a skyrmion in a magnetic multilayer disk at ~20 nm voxel resolution. This resolution is confirmed by Fourier shell correlation analysis. From the data the topological skyrmion number is computed directly, and the profile of the spin texture is probed at a length scale comparable to micromagnetic simulations. With direct vector operations on the data the spatial distributions of different energetic interactions are estimated and it is found that the DMI energy density is largest in the domain wall while the Heisenberg exchange energy density is more localized to the center of the structure. With another approach, the effects of altering energy terms on the topological charge profile shows that the DMI preferentially increases the size of a skyrmion smaller whereas the PMA decreases it. This is reflected in the data suggesting that the interactions in the sample are not uniform over the depth profile.

One intriguing possibility for future studies is measuring spin orbit coupling with combined spectra and microscopy to compute the spin orbit coupling interaction in three dimensions. This would allow the constants A, D, and K to be probed more directly from the data, rather than having to estimate them from the structure of the vector field. This would allow us to reconstruct the entire magnetic Hamiltonian from the magnetic state, which is the inverse to the typical problem of predicting the magnetic state from a Hamiltonian.

This work establishes that soft x-ray laminography using STXM has reached a level of maturity to probe magnetic stems at the same level as ubiquitous micromagnetic simulations with quantitative measurements of topological features which vary across depth. This level of sophistication in measurement and analysis will be necessary to probe the configuration space of material design for more complex three-dimensional textures such as Hopfions. Our results provide the foundation for nanoscale magnetic metrology for future tailored spintronics devices using topology as a design parameter, highlights the potential to reverse-engineer a spin Hamiltonian from macroscopic data, tying theory more closely to experiment.

## Materials and Methods

- **Thin film synthesis and nanopatterning**

The samples were grown by DC magnetron sputtering onto a 100nm thin SiNx membrane. The sample of interest was prepared in a tri-layer stack of 10X{Ir (1nm)/Co(1 nm)/Pt (1 nm)} / 10X{Ir (1nm)/Co(1.5 nm)/Pt (1 nm)}/ 10X{Ir (1nm)/Co(1 nm)/Pt (1 nm)} leading to a total thickness 95 nm (Fig. 1 b), which was chosen in view of a suitable X-ray absorption for soft X-rays around a photon energy of 778eV. Characterization with magnetic force microscopy (MFM) shows the well known multidomain structures in the variable Co thickness film (Fig. 1 a). Previous studies in similar multilayers have observed target skyrmions and Hopfions [3, 11]. For the x-ray laminography experiments, the multilayers were patterned into disks with diameters ranging from 200 nm to 1000 nm by electron beam lithography followed by lift-off.

- **3D tomography with soft X-ray laminography**

The soft x-ray magnetic laminography experiments were conducted at the PolLux beamline of the Swiss Light Source at the Paul Scherrer Institute in Villigen/Switzerland. Mounting the sample at 45 degrees incident angle to the x-ray beam a rotation series was taken over 56 rotation angles covering 360 degrees (Fig. 1 d-g). Three laminography data sets were recorded at the $L_3$ edge of Co around 778 eV x-ray photon energy with both positive and negative circular polarization, and linear horizontal polarization yielding three images at each rotation angle. The two circular polarization images allow to quantitatively derive the x-ray magnetic circular dichroism (XMCD) signal, whereas the linearly-polarized X-ray image enables removing additional non-linear structural background signals not accounted for with the typical dichroism $I_{XMCD} = \frac{I_\uparrow - I_\downarrow}{I_\uparrow + I_\downarrow}$, as well as unwanted asymmetries in the degree of circular polarization of the beam produced by the PolLux bending magnet [22, 23]. The 56 images covering 360 degrees are used to reconstruct a three-dimensional vectorized rendering. An iterative solver first creates a reconstruction of the topography of the disk. The three-dimensional topographic information is then used in a second step of the reconstruction algorithm, from which the spatially-resolved magnetization vector field is obtained (Fig. 2 a). Additional details on the reconstruction algorithms used for magnetic laminography can be found in Refs. [14, 23]. Each vector component $m_x$, $m_y$, $m_z$ was reconstructed separately from the data and then combined to produce the full 3D vector field (Fig. 2 b). Data are rendered using the visualization software *ParaView*$^{TM}$.

**Fourier shell analysis**

To confirm that the data are resolved on a sufficiently short length scale to calculate topological quantities in a meaningful way Fourier shell correlation (FSC) analysis is applied to each component of the magnetization. The Fourier shell correlation method is an established method for determining three-dimensional resolution in the field of electron microscopy and has been adopted in x-ray microscopy [8]. The data are divided into two categories set by the projection angle: one set relates to even indexed projections, the other set relates to odd indexed projections. From these half datasets two three-dimensional reconstructions are created. Comparing the phase of these reconstructions in Fourier space $FSC(r) = \frac{\sum F_1(r) F_2(r)^*}{(\sum |F_1(r)|^2 \sum |F_2(r)|^2)^{\frac{1}{2}}}$ where $F_1$ is the Fourier transform of the first set of images and $F_2$

of the second with ∗ indicating the complex conjugate, summed over all points at a radius $r$ in Fourier space [24, 25]. Using half-bit criteria, sinusoidal features are resolved at length scales with periods of $m_x$ = 94 nm, $m_y$ = 104 nm, $m_z$ = 86 nm. This confirms that the three-dimensional resolution in our experiment is sufficient on the length scale of the 95 nm thickness of our sample (Fig. 2 c).

**Acknowledgements**

This work was funded by the U.S. Department of Energy, Office of Science, Office of Basic Energy Sciences, Materials Sciences and Engineering Division under Contract No. DE-AC02-05-CH11231 (NEMM program MSMAG).

Work at the Molecular Foundry was supported by the Office of Science, Office of Basic Energy Sciences, of the U.S. Department of Energy under Contract No. DE-AC02-05CH11231.

The multilayer deposition was done at Western Digital Corp. San Jose CA under an MTA with LBNL (Berkeley Lab Ref. No. 2022-1213).

Part of this work was performed at the PolLux (X07DA) beamline of the Swiss Light Source, Paul Scherrer Institute, Villigen PSI, Switzerland. The PolLux endstation was financed by the German Bundesministerium für Bildung und Forschung through  contracts 05K16WED and 05K19WE2.

# Figures

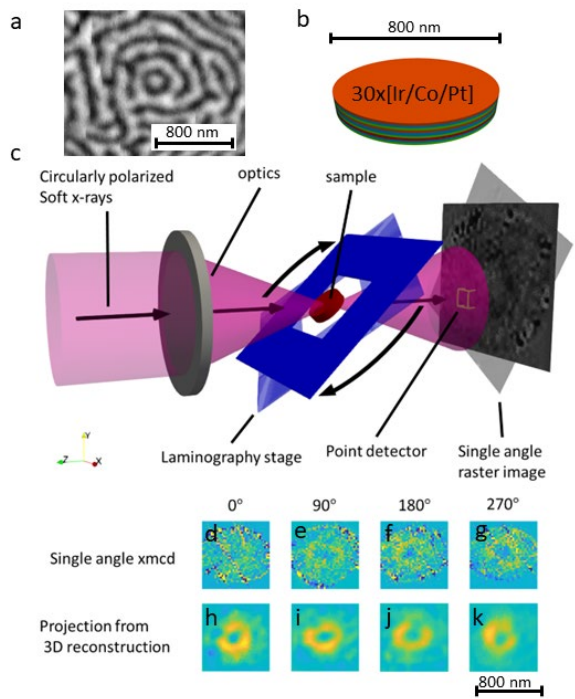

Fig. 1

a) Magnetic force microscopy phase shift image of an un-patterned film with the same composition as the disk used for the laminography reconstruction. A topological target skyrmion is visible in the MFM image.

b) Scale schematic of the magnetic multilayer disk used in the x-ray experiment. The disk was deposited on a SiNx membrane.

c) A schematic of the Pollux laminography beamline at the Swiss Light Source. The sample is mounted on a rotary stage for the measurements.

d-g) Pre-reconstruction single XMCD images calculated from individual projection angles data. The agreement between single data projections and reconstructed projections indicated good performance of the iterative reconstruction algorithm.

h-k) Post-reconstruction single XMCD images calculated from individual projection angles through the reconstruction. The full stack is used to create the 3D rendering.

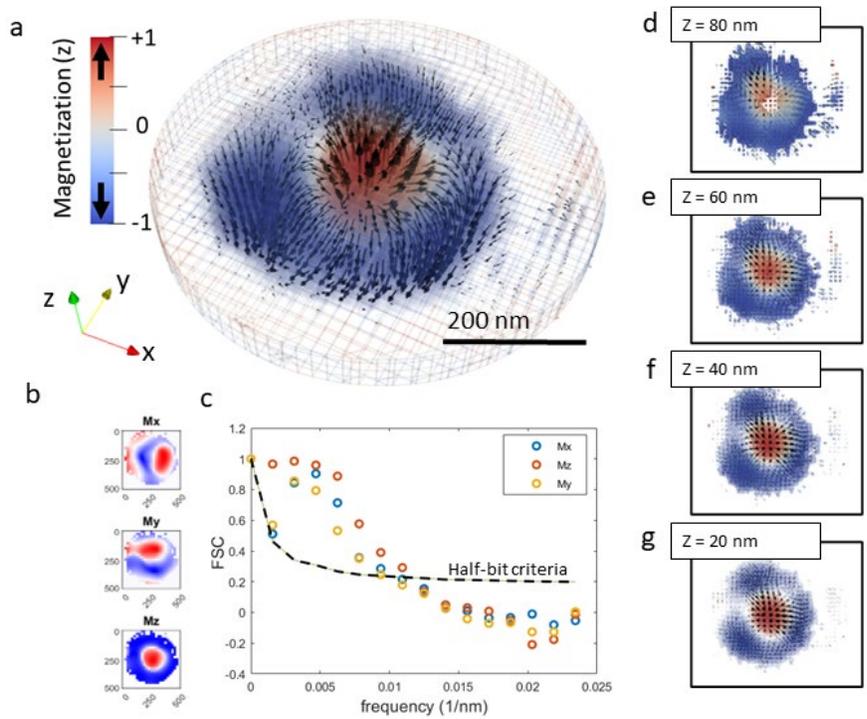

Fig. 2

a) A rendering in paraview of the product of the 3D reconstruction. The region of high statistical confidence contrast is ~500nm of the full 800nm diameter of the disk.. Blue-red color scaling maps the z-component of magnetization with red being up and blue being down.

b) Z-axis projections of individual cartesian magnetization components.

c) Fourier shell correlation analysis establishing the resolution at the intersection of the half bit curve with the FSC curve for each magnetization component.

d-g) Individual 20nm resolved slices along z-axis showing depth evolution of the skyrmion.

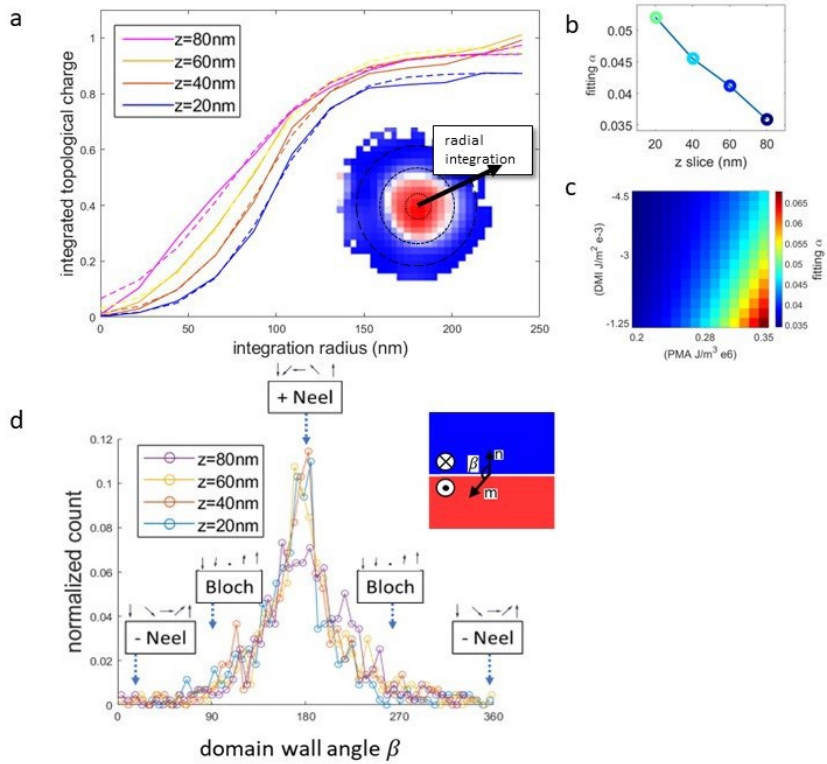

Fig. 3

a) Radially integrated topological charge profile across each depth in the reconstruction. Each curve is from a 20 nm thick voxel slice of the reconstruction in z. Data is solid line, fit is dashed line.

b) Fitting parameter $\propto$ as a function of z-position indicates a depth-dependent evolution of the spin texture's topological profile.

c) Micromagnetic simulations of a magnetic skyrmion spanning a phase space of DMI and PMA with the fitting parameter $\propto$ plotted as a color scale. Simulations indicate that changes in micromagnetic energy terms may be responsible for the depth dependent changes in fitting parameter.

d) Layer-resolved histogram of the chirality of the domain wall. The measured quantity at each voxel is the angle $\beta$ between the domain wall normal and the in-plane component of the magnetization. Chirality is constant across the z-profile indicating a positive chirality valued Neel wall. The inset illustration shows chirality angle $\beta$ of domain wall. The blue region represents a domain magnetized in -z and the red region represents a domain magnetized in +z. The vector n is the normal vector, m is the in-plane magnetization.

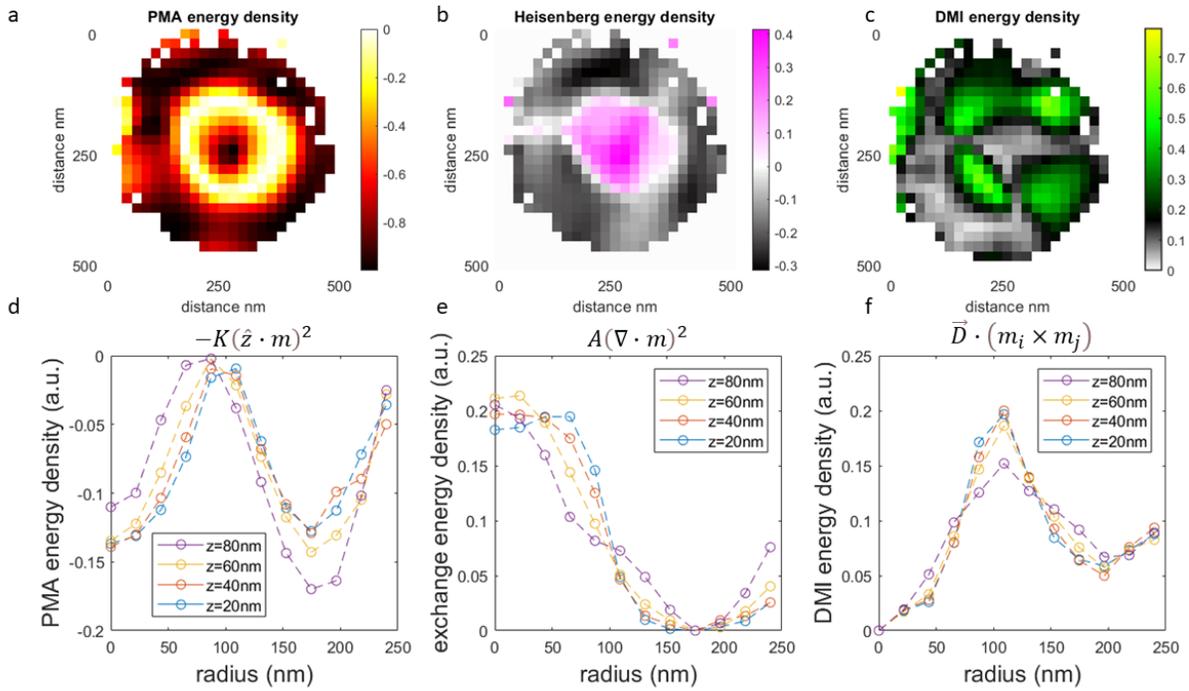

Fig. 4

a) Z-averaged PMA energy image with radial average for each individual 20 nm thick vertical slice. The energy is determined at the level of a relative normalized spatial density of the z-component squared of the magnetization, $-K(\hat{z}\cdot m)^2$.

b) Z-averaged exchange energy image with radial average for each individual 20 nm thick vertical slice. The energy is determined at the level of a relative normalized spatial density of the divergence squared of the magnetization, $A(\nabla \cdot m)^2$.

c) Z-averaged DMI energy image with radial average for each individual 20 nm thick vertical slice. The energy is determined at the level of a relative normalized spatial density of the interfacial chiral type ordering, $\vec{D}\cdot(m_i \times m_j)$ where $\vec{D}$ is a unit vector perpendicular to the displacement between spin cells, lying in-plane.

d) Radially averaged, normalized to area under curve PMA energy density profile (a.u).

e) Radially averaged, normalized to area under curve exchange interaction energy density profile (a.u).

f) Radially averaged, normalized to area under curve DMI energy density profile (a.u).